\documentclass[12pt]{article}
\usepackage{graphicx}
\usepackage{hyperref}

\def \bea{\begin{eqnarray}}
\def \beq{\begin{equation}}
\def \ca{{\cal A}}

\def \eea{\end{eqnarray}}
\def \eeq{\end{equation}}
\def \ko{K^0}

\def \ok{\overline{K}^0}

\def \s{\sqrt{2}}

\def \i{{\it i}}
\def \G{\Gamma}
\def \ovD{\overline{D}}
\def \of{\overline{f}}

\begin{document}

\rightline{EFI 12-1}
\rightline{UdeM-GPP-TH-12-205}
\rightline{TECHNION-PH-12-1}
\rightline{arXiv:1201.2351}
\rightline{January 2012}

\bigskip
\centerline{\bf CP ASYMMETRIES IN SINGLY-CABIBBO-SUPPRESSED}
\centerline{\bf $D$ DECAYS TO TWO PSEUDOSCALAR MESONS}
\bigskip
\centerline{Bhubanjyoti Bhattacharya}
\centerline{\it Physique des Particules, Universit\'e de Montr\'eal}
\centerline{\it C.P. 6128, succ.\ centre-ville, Montr\'eal, QC, Canada H3C 3J7}
\medskip

\centerline{Michael Gronau}
\centerline{\it Physics Department, Technion -- Israel Institute of Technology}
\centerline{\it Haifa 3200, Israel}
\medskip

\centerline{Jonathan L. Rosner}
\centerline{\it Enrico Fermi Institute and Department of Physics}
\centerline{\it University of Chicago, 5620 S. Ellis Avenue, Chicago, IL 60637}
\bigskip

\begin{quote}
The LHCb Collaboration has recently reported evidence for a CP asymmetry
approaching the percent level in the difference between $D^0 \to \pi^+ \pi^-$
and $D^0 \to K^+ K^-$.  We analyze this effect as if it is due to a penguin
amplitude with the weak phase of the standard model $c \to b \to u$ loop
diagram, but with a CP-conserving enhancement as if due to the
strong interactions.  In such a case the magnitude and strong phase of this
amplitude $P_b$ are correlated in order to fit the observed CP asymmetry, and
one may predict CP asymmetries for a number of other singly-Cabibbo-suppressed
decays of charmed mesons to a pair of pseudoscalar mesons.
Non-zero CP asymmetries are expected for $D^+ \to K^+ \ok$ (the most promising 
channel for which a 
non-zero CP asymmetry has not yet been reported), as well as $D^0 \to \pi^0
\pi^0$, $D_s^+ \to \pi^+ \ko$, and $D_s \to \pi^0 K^+$.  No CP asymmetry is
predicted for $D^+ \to \pi^+ \pi^0$ or $D^0 \to \ko \ok$ in this framework.
\end{quote}

\leftline{PACS numbers: 13.25.Ft, 11.30.Er, 11.30.Hv, 14.40.Lb}
\bigskip

\section{Introduction}
Although CP violation was first observed in neutral kaon decays and CP
asymmetries have been seen at the tens of percent level in $B$ meson decays,
the standard model describing these decays predicts
naturally very small CP asymmetries
in decays of charmed particles, of order $10^{-3}$ or less \cite{Bigi:2011re,%
Isidori:2011qw,Brod:2011re}.  These decays are dominated by physics of the
first two quark families, with the contribution of the third family suppressed
both by smallness of elements of the Cabibbo-Kobayashi-Maskawa (CKM)
matrix and by the relatively small $b$ quark mass in the $c \to b \to u$
penguin diagram.  This is in contrast to the $b \to t \to s$ penguin
amplitude, which can profit from both a larger CKM factor and a much larger
top quark mass.
Following an early suggestion~\cite{Golden:1989qx} that the penguin amplitude
in $D$ decays may be enhanced by nonperturbative effects in analogy to the
$s \to d$ penguin amplitude in $K \to \pi \pi$, recent studies
\cite{Isidori:2011qw,Brod:2011re,Bigi:2011em} indicate that an
order of magnitude enhancement is not impossible.

The LHCb Collaboration has reported $3.5 \sigma$ evidence for CP-violating
charm decays in the difference between CP asymmetries in $D^0 \to K^- K^+$
and $D^0 \to \pi^- \pi^+$
\cite{Aaij:2011in}:
\beq \label{eqn:LHCb}
\Delta A_{CP} \equiv A_{CP}(K^+ K^-) - A_{CP}(\pi^+ \pi^-) = [-0.82
\pm 0.21({\rm stat}) \pm 0.11({\rm syst})]\%~.
\eeq
For the decay of a charmed meson $D$ to any final state $f$ we are defining
\beq
A_{CP}(f) \equiv \frac{\Gamma(D \to f) - \Gamma(\bar D \to \bar f)}
                      {\Gamma(D \to f) + \Gamma(\bar D \to \bar f)}~.
\eeq
Although the CDF II Collaboration at the Fermilab Tevatron does not see
statistically compelling evidence for CP violation in either of these two
decays, their results are consistent with those of LHCb \cite{Aaltonen:2011se}:
\beq
A_{CP}(D^0 \to K^+ K^-) = (-0.24 \pm 0.22 \pm 0.09)\%,~
A_{CP}(D^0 \to \pi^+ \pi^-) = (0.22 \pm 0.24 \pm 0.11)\%~.
\eeq
We calculate the corresponding 90\% confidence level limits to be
\beq \label{eqn:limits}
-0.63\% \le A_{CP}(D^0 \to K^+ K^-) \le 0.15\%~,~~
-0.21\% \le A_{CP}(D^0 \to \pi^+ \pi^-) \le 0.65\%~.
\eeq

The LHCb results have led to numerous hypotheses of physics beyond the
standard model (e.g.,
\cite{Isidori:2011qw,Wang:2011uu,Hochberg:2011ru,Rozanov:2011gj})
some of which had been studied earlier \cite{Grossman:2006jg}.
A more conservative approach, studying the above CP asymmetries within the
Standard Model under relaxed assumptions about non-perturbative hadronic weak
matrix elements, has been adopted recently in two papers applying flavor SU(3).
Ref.\ \cite{Pirtskhalava:2011va} extended the hypothesis of triplet operator
enhancement introduced in Ref.\ \cite{Golden:1989qx} by including in the
effective weak Hamiltonian SU(3) breaking terms which are first order in the
strange quark mass. A second work \cite{Cheng:2012wr} (appearing while we were
writing up our results), applying a diagramatic flavor SU(3) approach similar
to the one discussed by us below, associated $W$-exchange and annihilation
amplitudes with final state resonant effects \cite{Cheng:2010ry}.  While these
two papers have some overlap with ours the specific assumptions and detailed
predictions of the three studies, all based on flavor SU(3) analyses,
are different.

In the present paper we explore a scenario in which the standard model $c \to
b \to u$ penguin amplitude receives a sufficient enhancement from strong
interaction physics to account for the effect observed by LHCb.
This amplitude then must contribute to other direct
CP asymmetries in decays of $D^0$, $D^+$, and $D_s^+$ to pairs of
pseudoscalar mesons. Non-zero CP asymmetries are expected for $D^+ \to K^+
\ok$, as well as $D^0 \to \pi^0 \pi^0$, $D_s^+ \to \pi^+ \ko$, and $D_s \to
\pi^0 K^+$.  No CP asymmetry is predicted for $D^+ \to \pi^+ \pi^0$ or $D^0 \to
\ko \ok$ in this model.  The former receives no penguin contribution, being a
$\Delta I = 3/2$ process, while the latter involves a different penguin
amplitude than the one we are considering.

We perform the analysis in the context of a flavor-SU(3) model of charm decays
presented previously \cite{Bhattacharya:2008ke,Bhattacharya:2009ps}.
We introduce notation in Sec.\ II and fit decay rates for singly-Cabibbo-%
suppressed (SCS) processes (including SU(3) breaking) in Sec.\ III.  We then
introduce a phenomenological penguin amplitude $P_b$ in Sec. IV to account for
the CP violation observed by LHCb, and predict other CP asymmetries for SCS
charmed meson decays.  We summarize in Sec.\ V.

\section{Formalism and notation}

\begin{figure}
\begin{center}
\vskip -0.5cm
\includegraphics[width=0.45\textwidth]{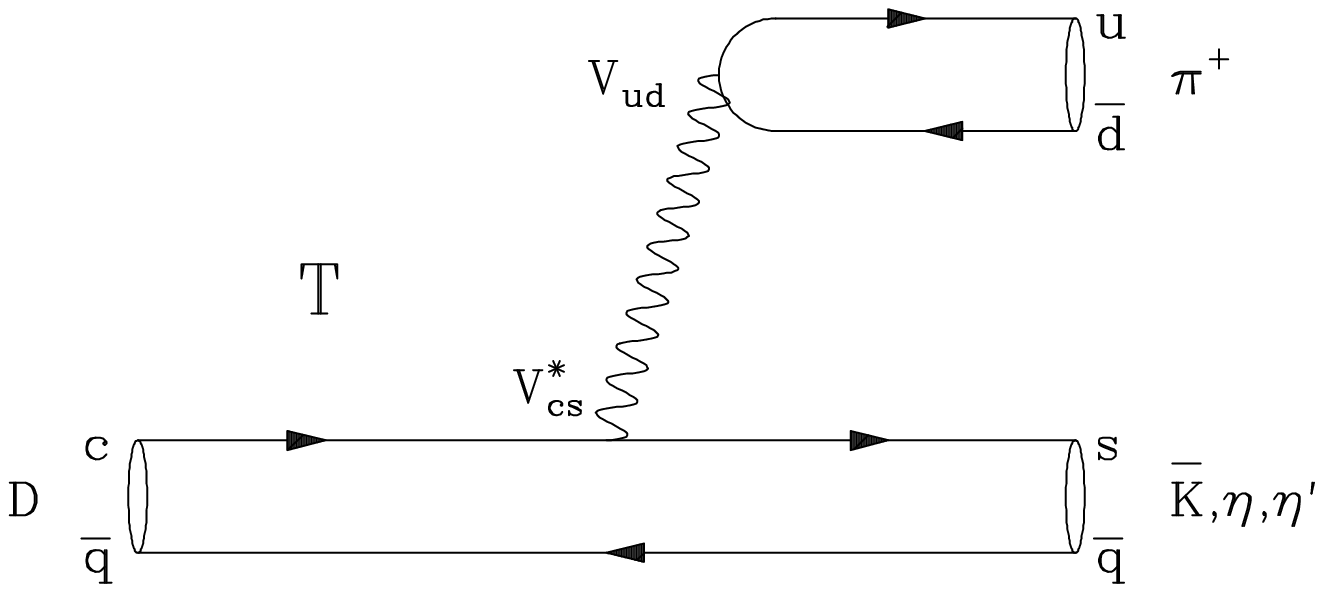} \hspace{0.5cm}
\includegraphics[width=0.45\textwidth]{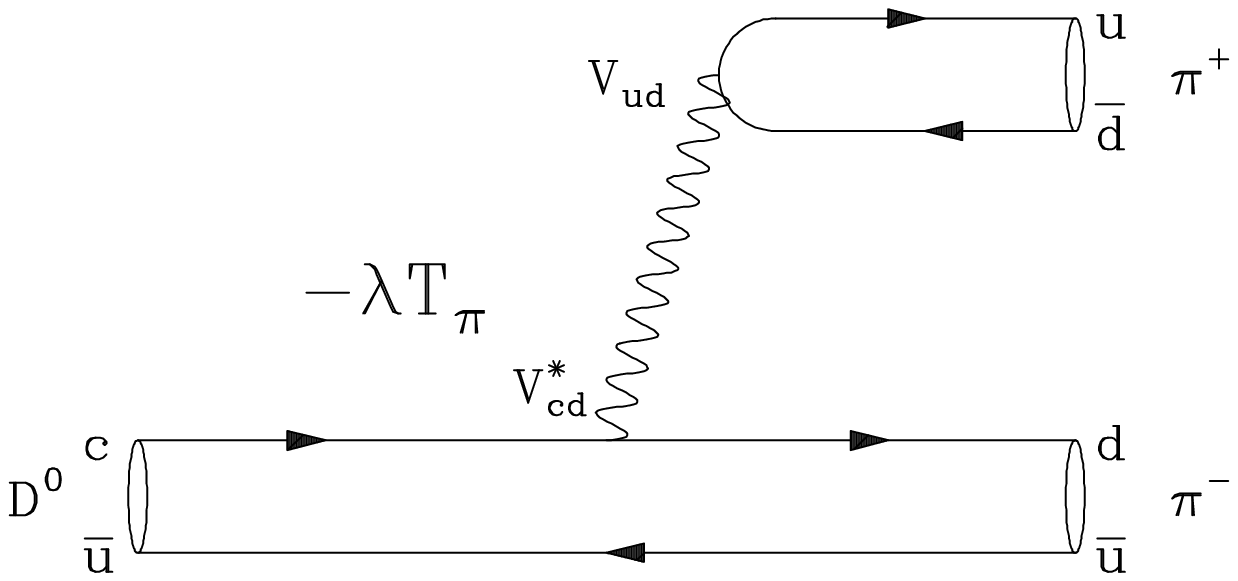}
\vskip 0.5cm
\includegraphics[width=0.45\textwidth]{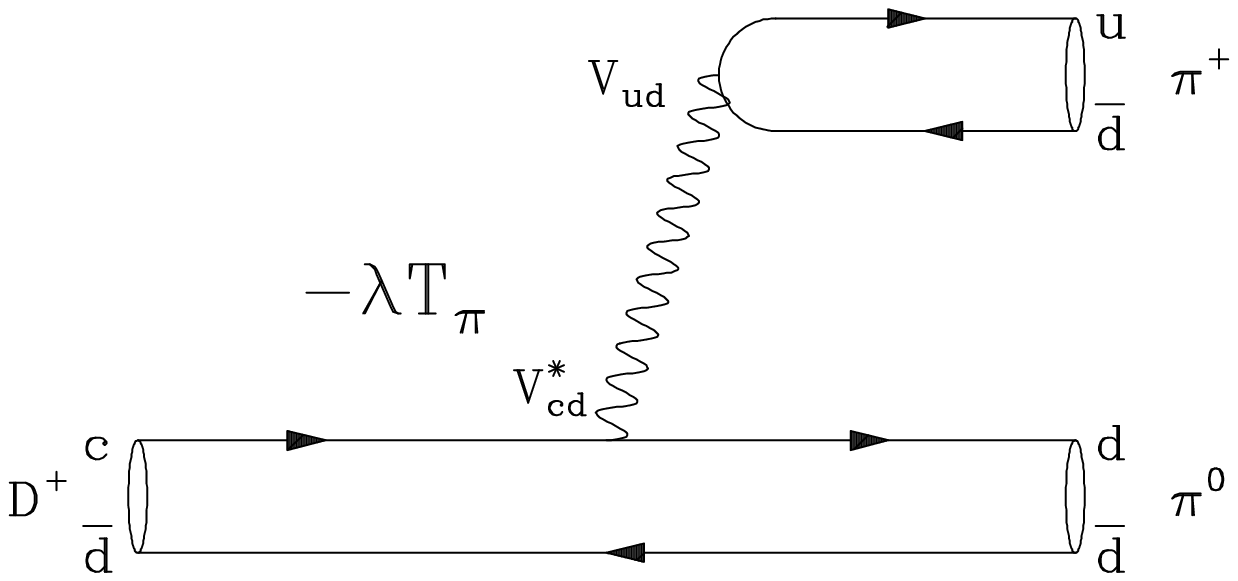} \hspace{0.5cm}
\includegraphics[width=0.45\textwidth]{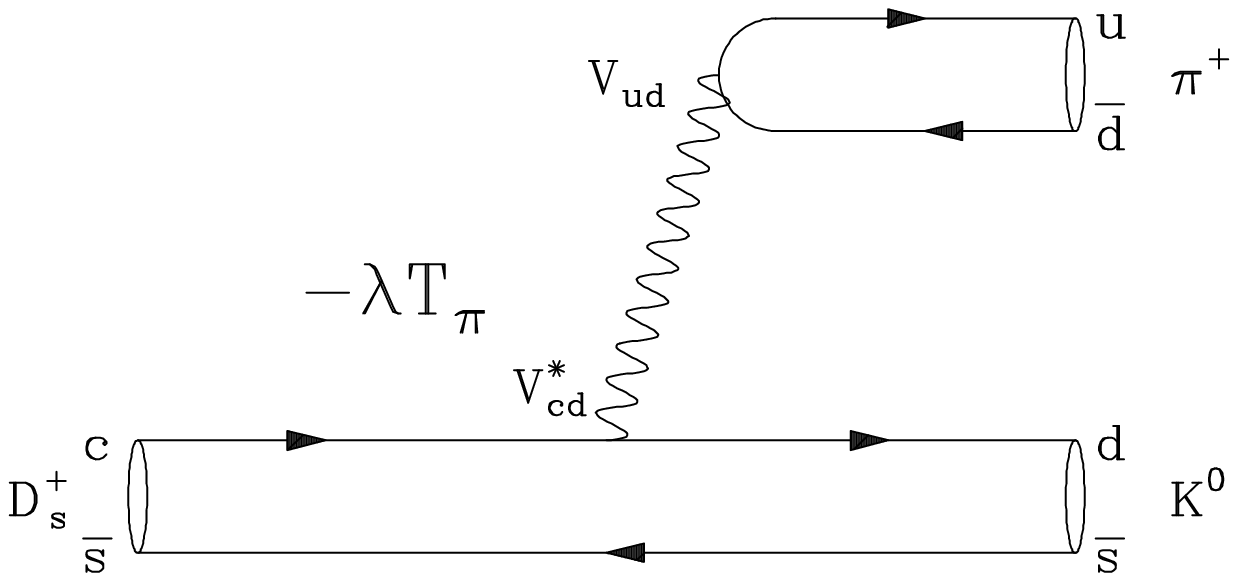}
\vskip 0.5cm
\includegraphics[width=0.45\textwidth]{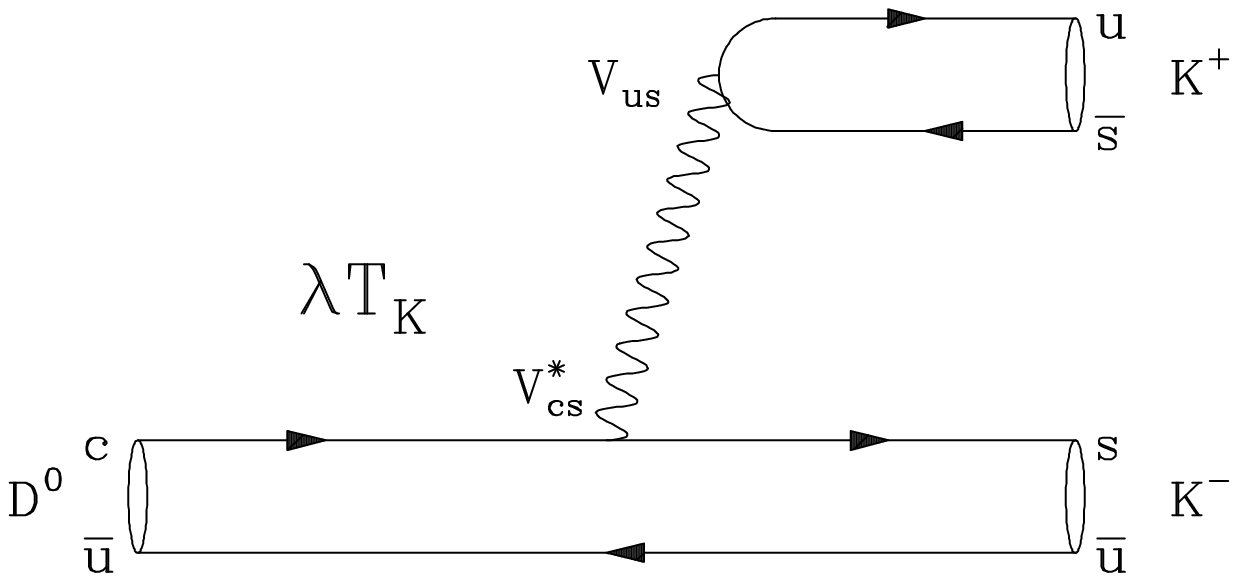} \hspace{0.5cm}
\includegraphics[width=0.45\textwidth]{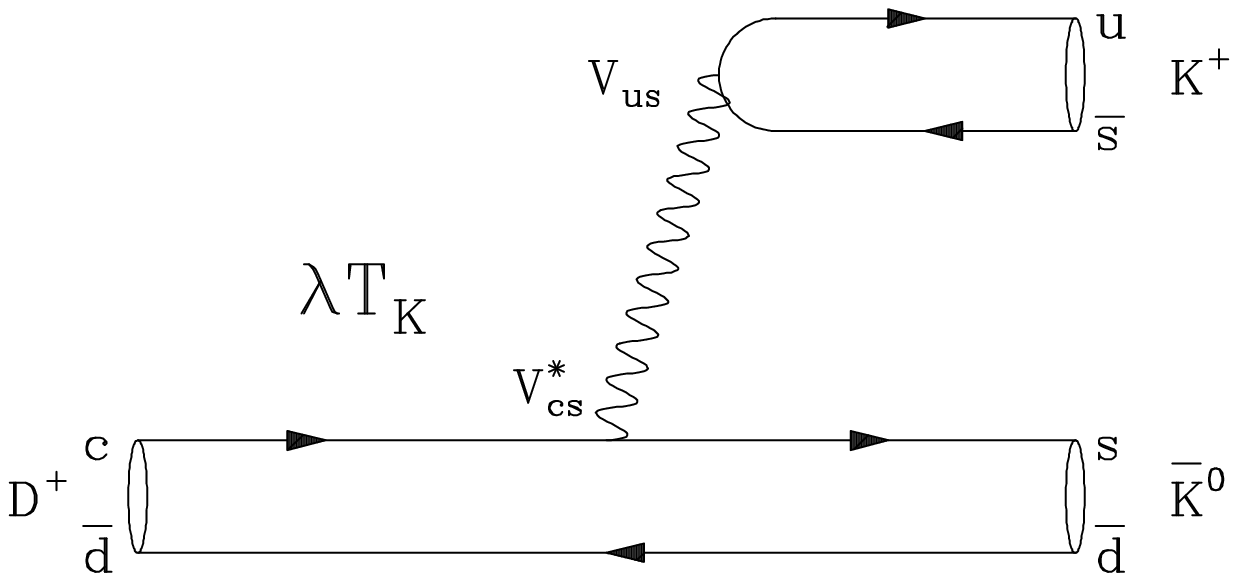}
\end{center}
\vskip -0.5cm
\caption{Relation between CF and SCS color-favored tree amplitudes.
\label{fig:T}}
\end{figure}

Cabibbo-favored (CF) charm decays may be characterized by amplitudes $T$, $C$,
$E$, and $A$, corresponding to color-favored tree, color-suppressed tree,
exchange, and annihilation flavor topologies \cite{Bhattacharya:2008ke,%
Bhattacharya:2009ps}.  A fit to CF decays of $D$ mesons to two pseudoscalar
mesons leads to the following ($|T| > |C|$) solution:
\bea
T &=& 2.927~, \\
C &=& 2.337\,e^{-\,\i\,151.66^\circ} = -2.057 - 1.109~i~, \label{C}\\
E &=& 1.573\,e^{~\i\,120.56^\circ} = -0.800 + 1.355~i~, \\
A &=& 0.33\,e^{~\i\,70.47^\circ} = 0.110 + 0.311~i~\label{E},
\eea
quoted in units of $10^{-6}$ GeV.  We note that complex conjugates of the
amplitudes (\ref{C})-(\ref{E}) give identical decay rates.  To describe
amplitudes corresponding to SCS processes the above amplitudes are multiplied
by $\pm \lambda$, where $\lambda = \tan \theta_{\rm Cabibbo} = 0.2317$.  (Tiny
phases of $V_{ud}$, $V_{cs}$, $V_{cd}$, and $V_{us}$ are neglected. These
contribute to negligible direct CP asymmetries in $D^0\to \pi^+\pi^-$ and $D^0
\to K^+K^-$ at a level below one per thousand by interference of the above
amplitudes with SU(3)-breaking penguin amplitudes discussed in Section III.) 
The relation between CF and SCS color-favored tree
amplitudes is illustrated in Fig.\ \ref{fig:T}.

In order to account for SU(3) breaking in the SCS $T$ amplitude we may use the
following expressions:
\bea
T_{D^0\to\pi^+\pi^-}~~=~~T_{D^+\to\pi^+\pi^0}~~=~~T_{D^+_s\to \pi^+ K^0}~~
=~~T_\pi~,\\
T_{D^0\to  K^+ K^-}~~=~~T_{D^+\to K^+ \ok}~~=~~T_K~,~~~~~~~~
\eea
where, neglecting the contribution of $f_-(q^2)$ at $q^2 = m^2_{\pi, K}$,
\bea
T_\pi &=& T\,\cdot\,\frac{|f_{+(D^0\to\pi^-)}(m^2_\pi)|}{|f_{+(D^0\to K^-)}
(m^2_\pi)|} \,\cdot\,\frac{m^2_D - m^2_\pi}{m^2_D - m^2_K}, \\
T_K   &=& T\,\cdot\,\frac{|f_{+(D^0\to  K^-)}(m^2_K)|}{|f_{+(D^0\to K^-)}
(m^2_\pi)|} \,\cdot\,\frac{f_K}{f_\pi}.
\eea

\begin{figure}
\begin{center}
\vskip -0.5cm
\includegraphics[width=0.45\textwidth]{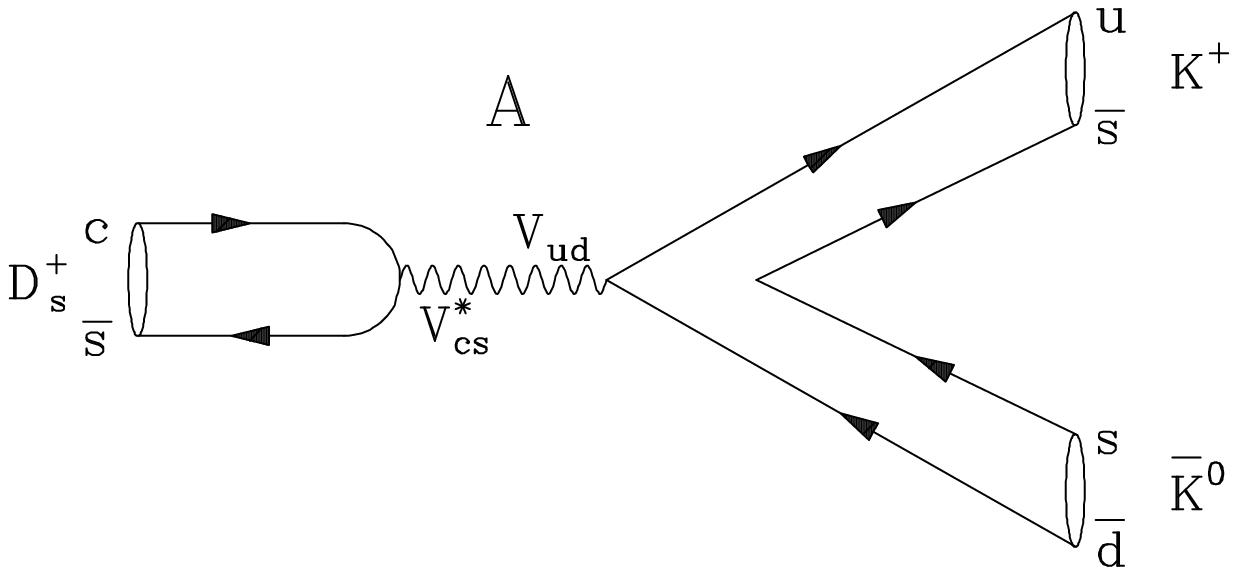} \hspace{0.5cm}
\includegraphics[width=0.45\textwidth]{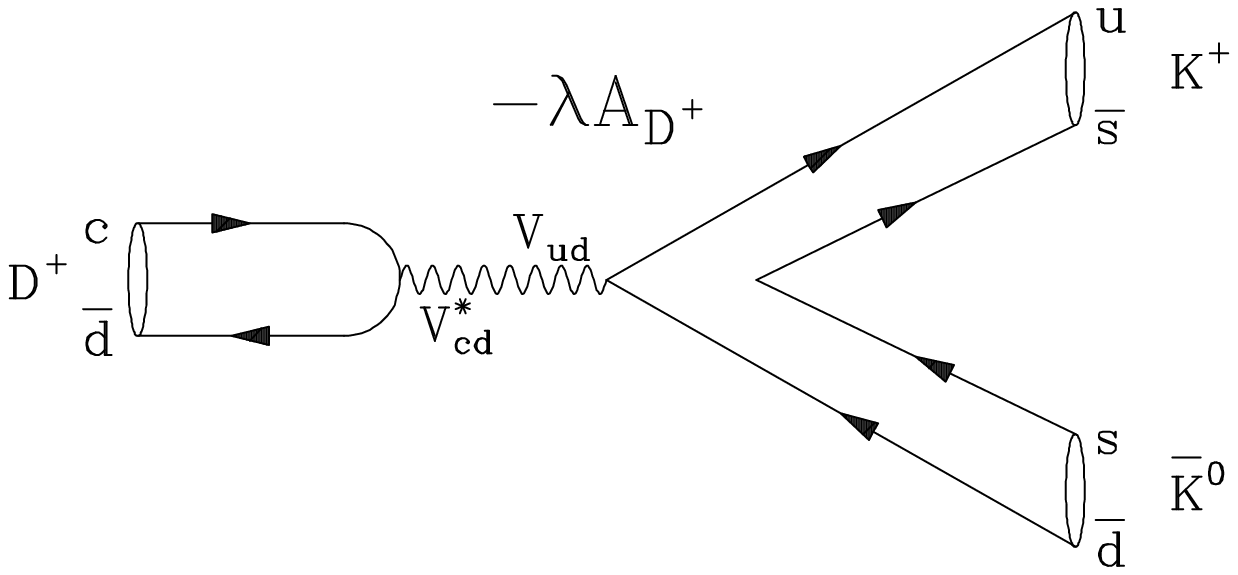}
\vskip 0.5cm
\includegraphics[width=0.45\textwidth]{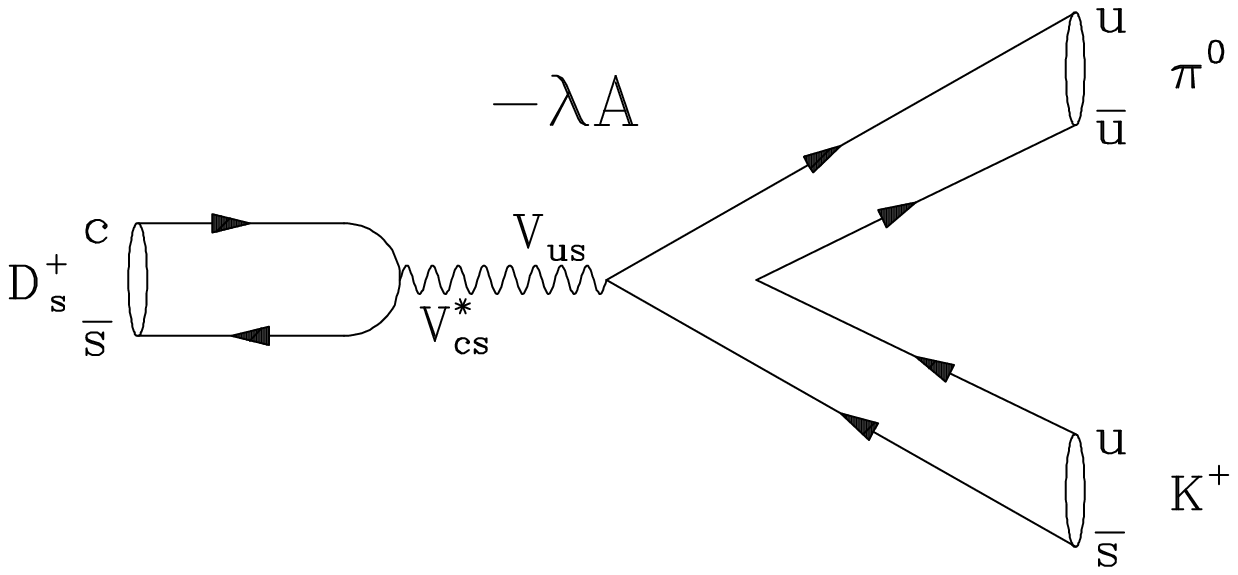} \hspace{0.5cm}
\includegraphics[width=0.45\textwidth]{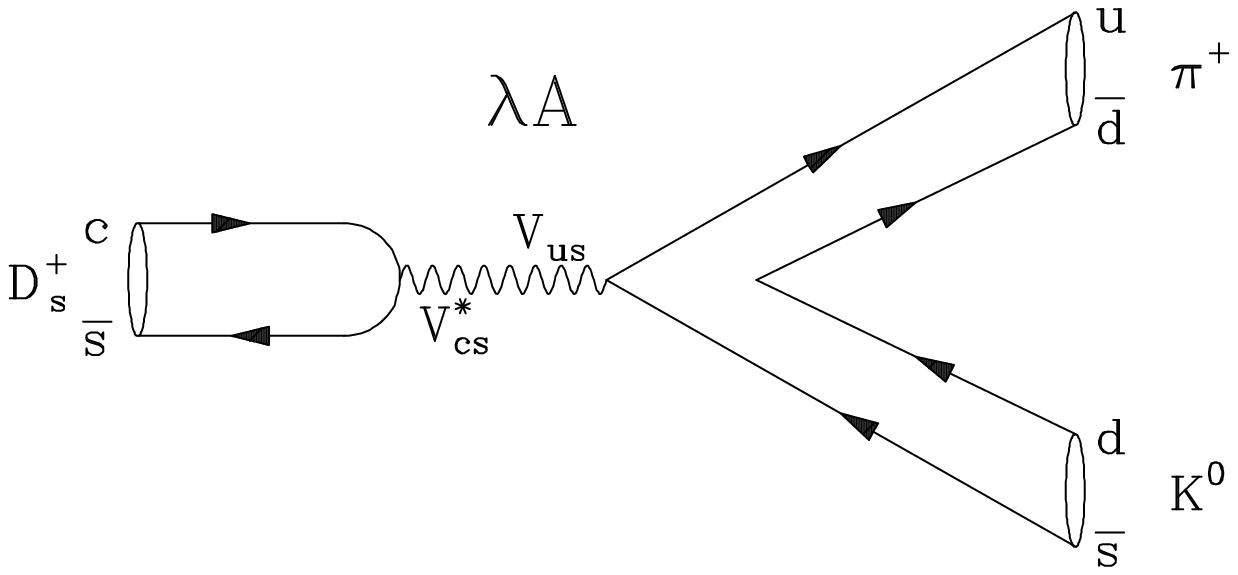}
\vskip 0.5cm
\end{center}
\caption{Relation between CF and SCS $A$ amplitudes.
\label{fig:A}}
\end{figure}

Similarly, the relation between
CF and SCS $A$ amplitudes is illustrated in
Fig.\ \ref{fig:A}.  Here, one may introduce SU(3) breaking as follows:
\bea
A_{D^+\to K^+\ok} &=& A\,\cdot\,\frac{f_{D^+}}{f_{D^+_s}}~~=~~A_{D^+}~, \\
A_{D^+_s\to\pi^+ K^0} &\simeq& A_{D^+_s\to K^+\pi^0}~~=~~A~.
\eea
We know the relevant decay constants \cite{Rosner:2010} and meson masses
\cite{Nakamura:2010zzi} (in GeV):
\bea
f_\pi = 0.13041;~~~f_K = 0.1561;~~~f_{D^+} = 0.2067;~~~f_{D^+_s} = 0.2575;\\
m_{D^0} = 1.8648;~~~m_\pi = 0.13957018;~~~m_K = 0.493677.~~~~~~
\eea
The following approximate values are also known for the form factors from
semileptonic $D^0$ decays \cite{Shipsey:2007, Besson:2009uv}:
\bea
|f_{+(D^0\to\pi^-)}(m^2_\pi)| \simeq 0.705~, \\
|f_{+(D^0\to  K^-)}(m^2_\pi)| \simeq 0.768~, \\
|f_{+(D^0\to  K^-)}(m^2_K)  | \simeq 0.811~.
\eea
After using relevant form factors and decay constants, we find, in units of
$10^{-6}$ GeV,
\bea
T_\pi &=& 2.87~, \\
T_K  &=& 3.70~, \\
A_{D^+} &=& (0.89 + 2.50 \, \i) \times 10^{-1}~.
\eea
The tree-level amplitudes for $D^0 \to \pi^+ \pi^-$, $D^0 \to K^+ K^-$,
and $D^0 \to \pi^0 \pi^0$ involve the following respective combinations
(in units of $10^{-7}$ GeV):
\bea
-\lambda\,(T_\pi + E) &=& - 4.80 - 3.14\, \i~, \label{eqn:pipi}\\
 \lambda\,(T_K + E)  &=& ~~ 6.72 + 3.14\, \i~, \label{eqn:KK}\\
 \lambda\,(C - E) &=& -2.91 - 5.71\, \i~, \label{eqn:pizpiz}
\eea
as well as SU(3)-breaking terms which we shall now introduce.

\section{Fits to decay rates including SU(3) violation}

\begin{figure}
\begin{center}
\vskip -0.5cm
\includegraphics[width=0.45\textwidth]{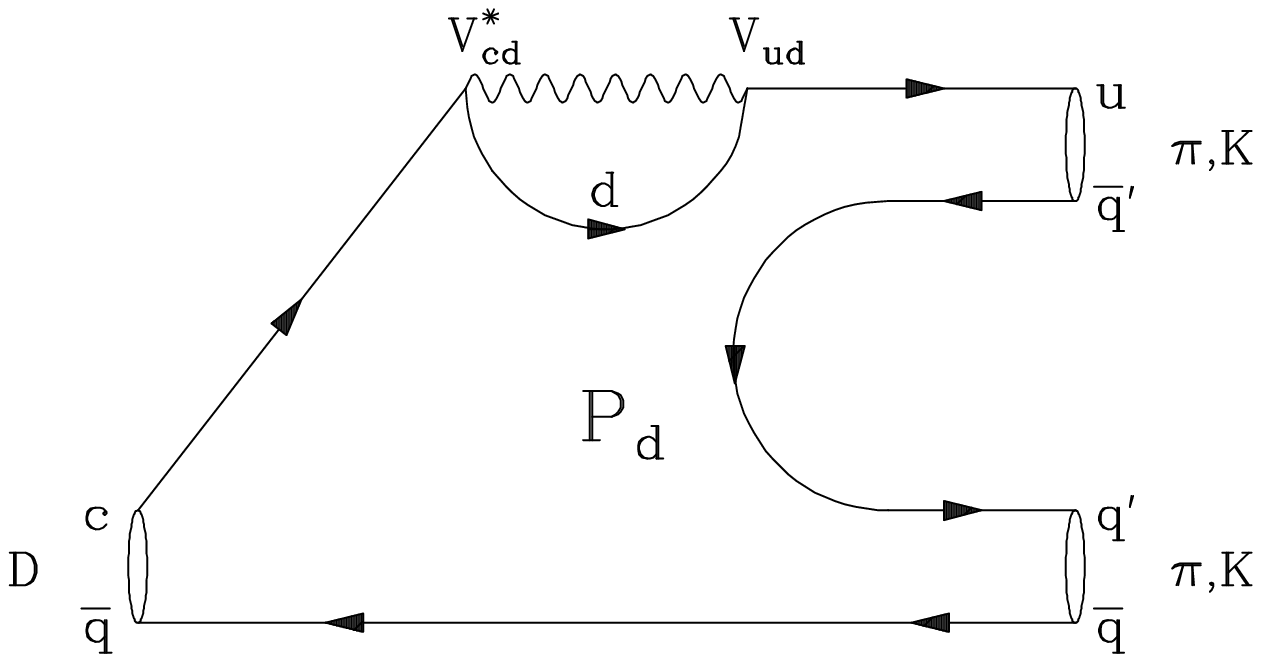} \hspace{0.5cm}
\includegraphics[width=0.45\textwidth]{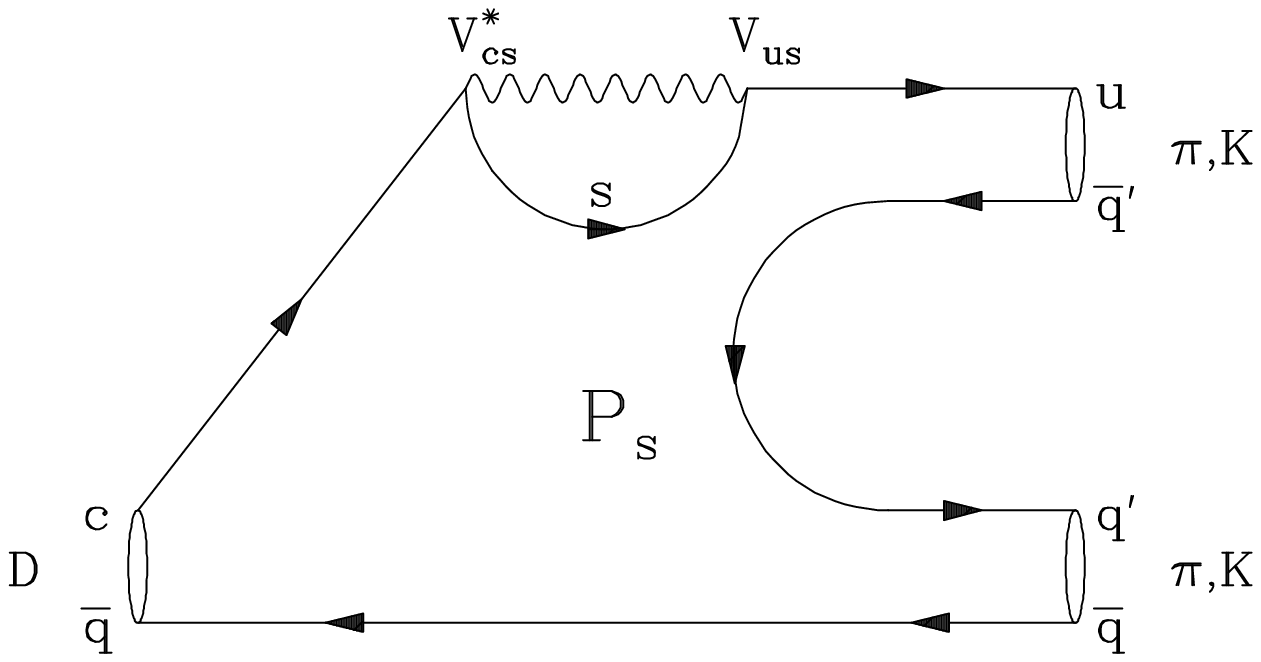}
\vskip 0.5cm
\end{center}
\caption{Penguin diagrams leading to a non-zero amplitude $P = P_d + P_s$ in
the presence of imperfect cancellation between intermediate $d$ and $s$ quarks.
\label{fig:P}}
\end{figure}

\begin{figure}
\begin{center}
\vskip 0.5cm
\includegraphics[width=0.45\textwidth]{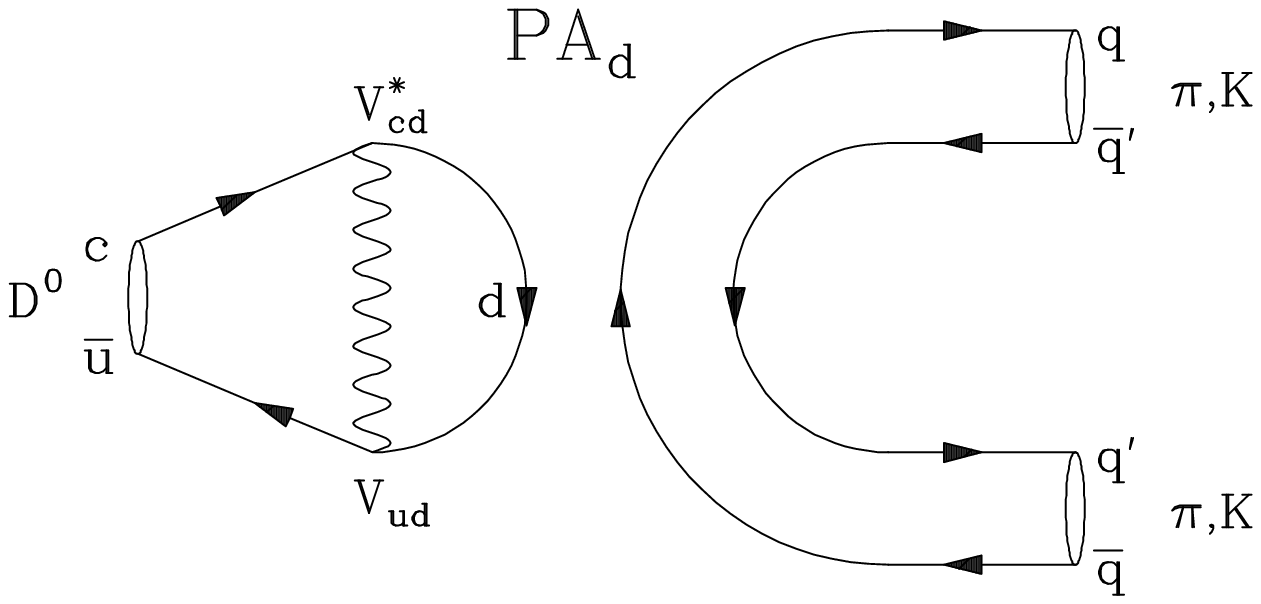} \hspace{0.5cm}
\includegraphics[width=0.45\textwidth]{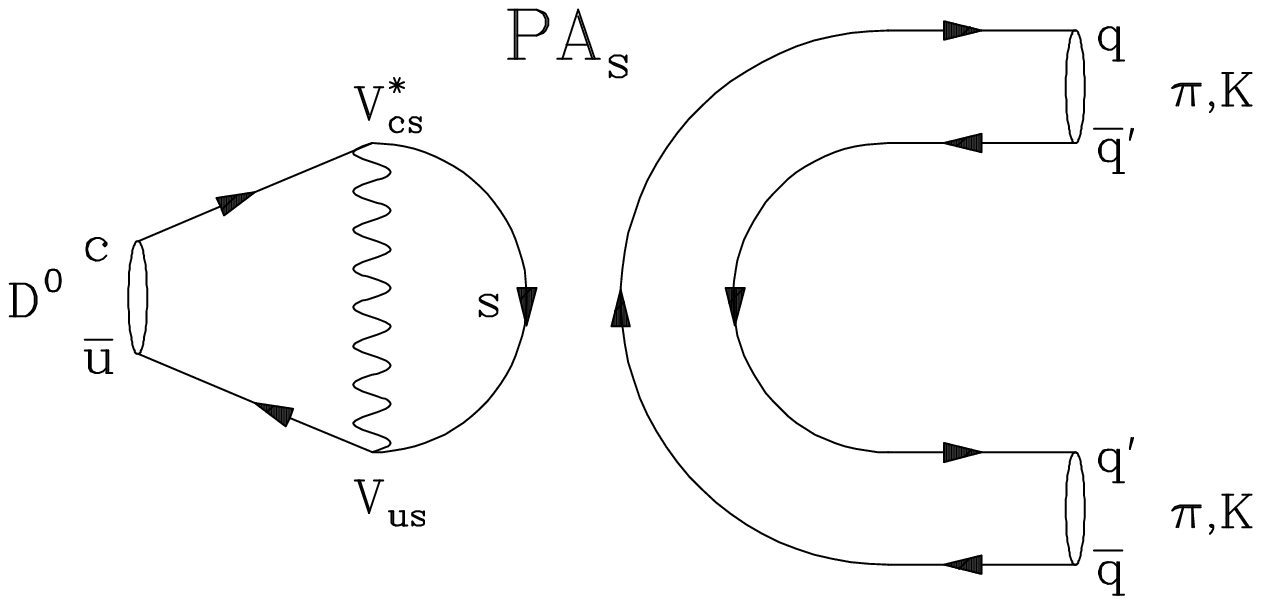}
\vskip 0.5cm
\end{center}
\caption{Penguin annihilation diagrams leading to a non-zero amplitude $PA =
PA_d + PA_s$ in the presence of imperfect cancellation between intermediate
$d$ and $s$ quarks.
\label{fig:PA}}
\end{figure}

The penguin amplitude $P$ for $c \to u$ transitions is normally thought to be
very small because the contributions of $d$ and $s$ quarks in the intermediate
state cancel one another \cite{Glashow:1970gm}.  If we regard this cancellation
as inexact due to SU(3)-violating masses of intermediate-state particles, we
can regard the penguin amplitude $P$ as a proxy for SU(3) violation (see Fig.\
\ref{fig:P}).  It will then have the same weak phase as other standard model
contributions to $D$ decays.
The same can be said for a penguin annihilation ($PA$) amplitude, contributing
only to $D^0$ decays.  It corresponds to the exchange processes $c \bar u \to s
\bar s$ and $c \bar u \to d \bar d$ followed by $s \bar s$ or $d \bar d$
annihilation into a pair of charge-conjugate pseudoscalar mesons (Fig.\
\ref{fig:PA}).

\begin{figure}
\begin{center}
\includegraphics[width=0.48\textwidth]{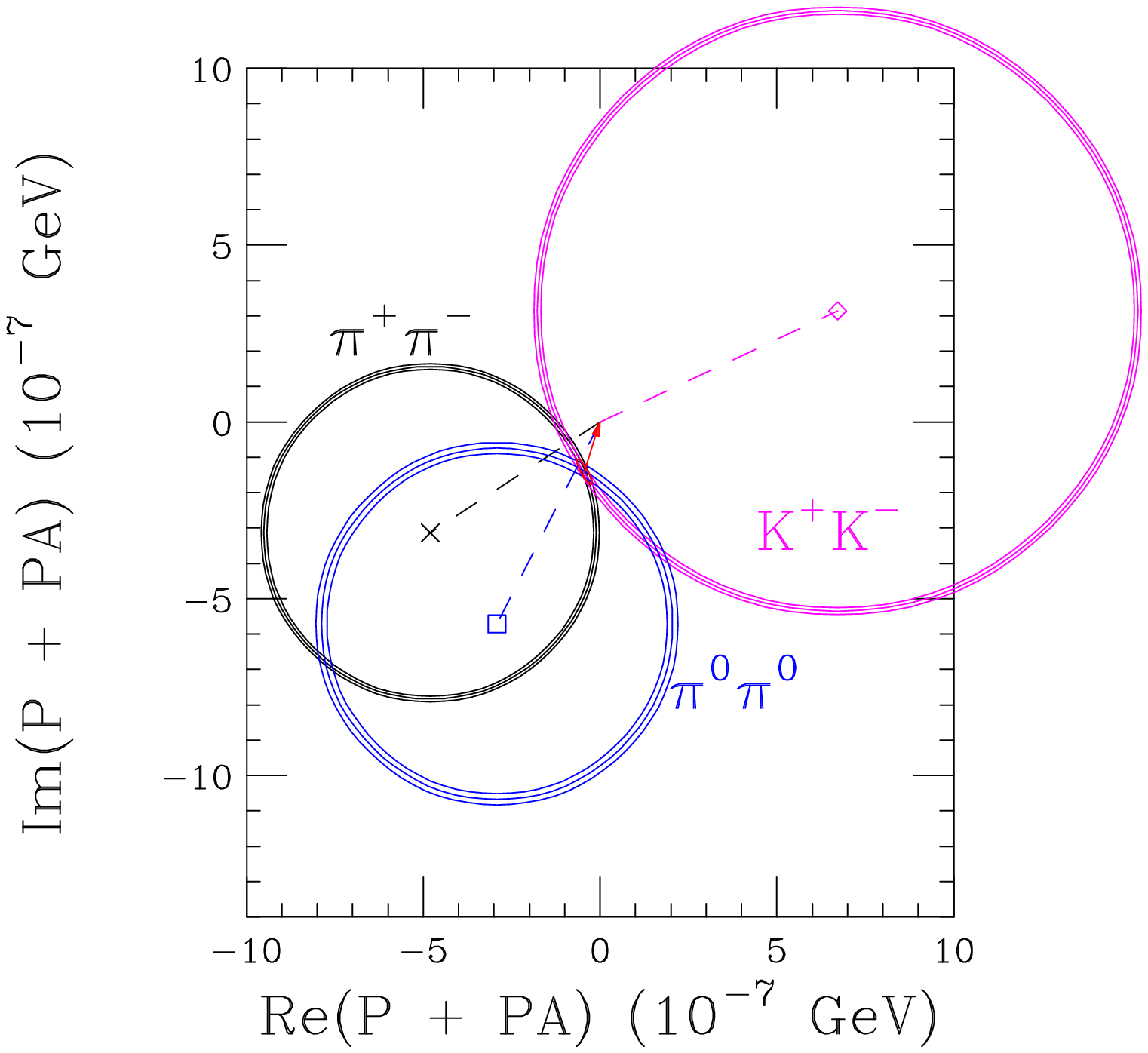} \hspace{0.4cm}
\includegraphics[width=0.42\textwidth]{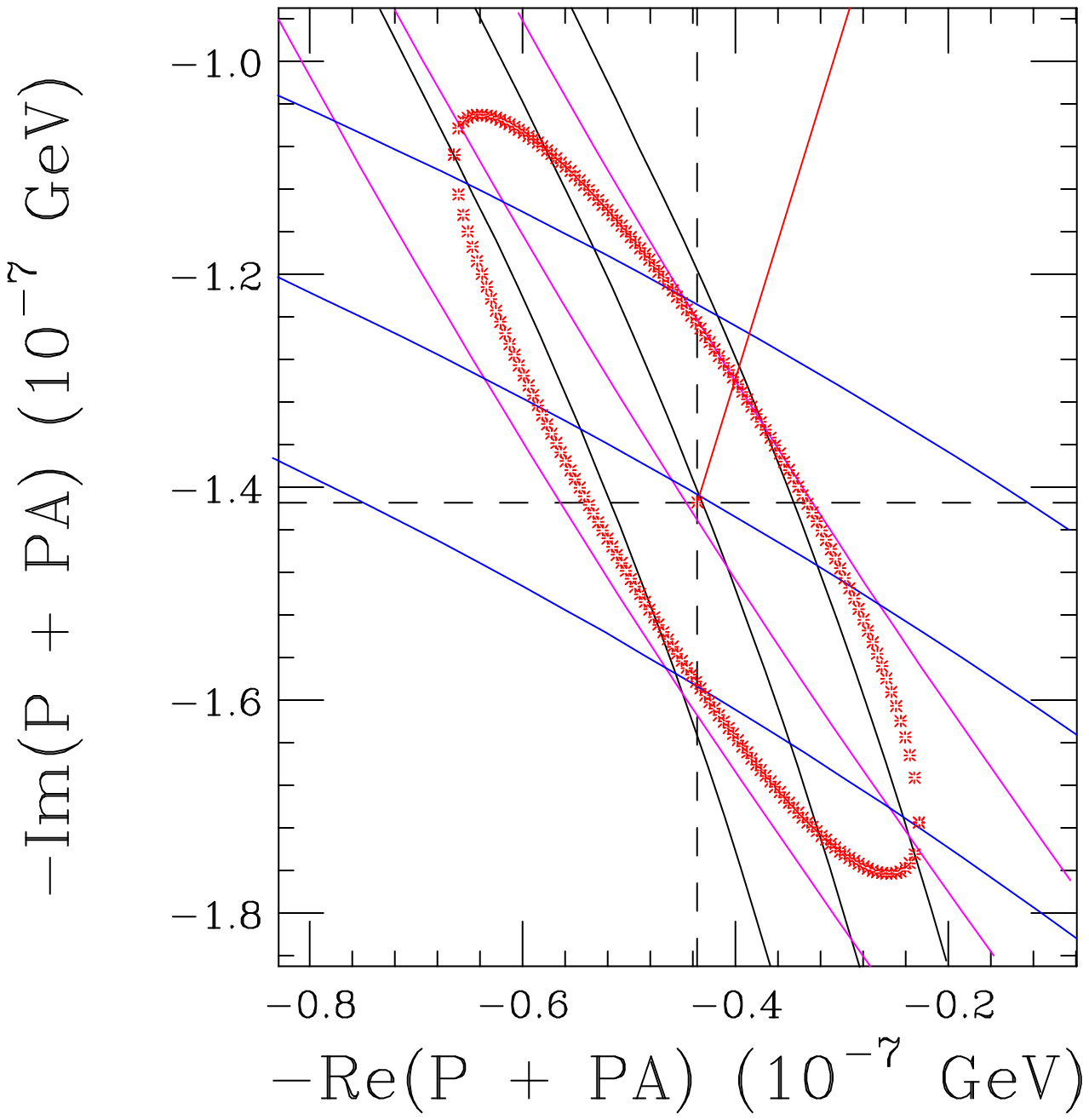}
\caption{Construction to determine $P + PA$.
The relative sign between the left-hand and (magnified) right-hand
panels is due to the fact that the vector $P + PA$ points {\it toward}
the origin in the left-hand figure.
 \label{fig:pen}}
\end{center}
\end{figure}

\begin{table}
\caption{Representations and comparison of experimental and fit
amplitudes for SCS decays of charmed mesons to two pseudoscalar mesons.
\label{tab:pen}}
\begin{center}
\begin{tabular}{c c c c c} \hline \hline
Decay & Amplitude      & \multicolumn{2}{c}{$|\ca|$ ($10^{-7}$ GeV)} & $\chi^2$ \\ \cline{3-4}
 Mode & representation & Experiment & Theory Fit \\ \hline
$D^0\to\pi^+\pi^-$ &$-\lambda\,(T_\pi + E) + (P + PA)$  &4.70$\pm$0.08&4.70&0\\
$D^0\to  K^+  K^-$ &~$\lambda\,(T_K + E) + (P + PA)$ &8.49$\pm$0.10&8.48&0.01\\
$D^0\to\pi^0\pi^0$ &$-\lambda\,(C - E)/\s - (P + PA)/\s$&3.51$\pm$0.11&3.51&0\\ \hline
$D^+\to\pi^+\pi^0$ &$-\lambda\,(T_\pi + C)/\s$ &2.66$\pm$0.07&2.26&33\\ \hline
$D^0\to K^0\ok$    &$-(P + PA) + P$                 &2.39$\pm$0.14&2.37&0.02\\
$D^+\to K^+\ok$    &~$\lambda\,(T_K - A_{D^+}) + P$    &6.55$\pm$0.12&6.87&7\\
$D^+_s\to\pi^+ K^0$&$-\lambda\,(T_\pi - A) + P$        &5.94$\pm$0.32&7.96&40\\
$D^+_s\to\pi^0 K^+$&$-\lambda\,(C + A)/\s - P/\s$      &2.94$\pm$0.55&4.44&7\\
\hline \hline
\end{tabular}
\end{center}
\end{table}
The amplitudes for $D^0 \to \pi^+ \pi^-$, $D^0 \to K^+ K^-$, and $D^0 \to
\pi^0 \pi^0$ then may be expressed as shown in the first three lines of
Table \ref{tab:pen}.  Given the magnitudes of the relevant amplitudes
determined from the decay rates \cite{Bhattacharya:2009ps, Amsler:2008zzb},
in units of $10^{-7}$ GeV,
\bea
|\ca(D^0\to\pi^+\pi^-)| &=& 4.70 \pm 0.08~, \\
|\ca(D^0\to K^+ K^-)|   &=& 8.49 \pm 0.10~, \\
\s\,|\ca(D^0\to\pi^0\pi^0)| &=& 4.96 \pm 0.16~,
\eea
one may then plot circles with these radii and centers defined by Eqs.\
(\ref{eqn:pipi}--\ref{eqn:pizpiz}) to solve for a common value of $P+PA$.
The existence of a self-consistent solution for $P+PA$ is supported by a
$\chi^2$--minimization fit, which leads to
\beq
P + PA = [(0.44\pm0.23) + (1.41\pm0.36)~i] \times 10^{-7}~{\rm GeV}~;~~~~~~~
\chi^2/{\rm d.o.f.} = 0.012/1 = 0.012~.
\eeq
The construction and the corresponding $\Delta \chi^2 = 2.3$ error ellipse
(corresponding to 68\% probability) are shown in Fig.\ \ref{fig:pen}.

%
\begin{figure}
\begin{center}
\includegraphics[width=0.5\textwidth]{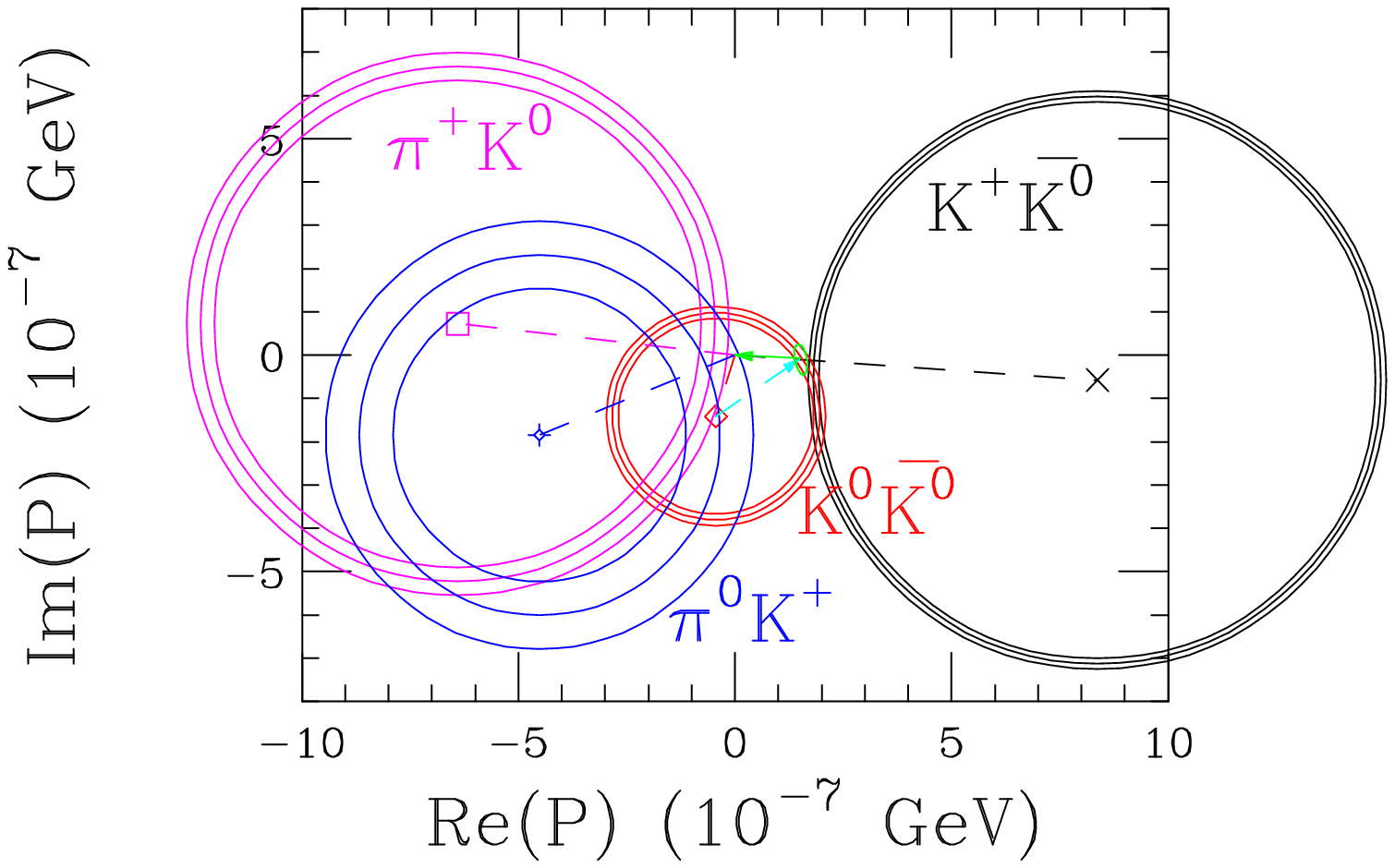} \hspace{0.2cm}
\includegraphics[width=0.44\textwidth]{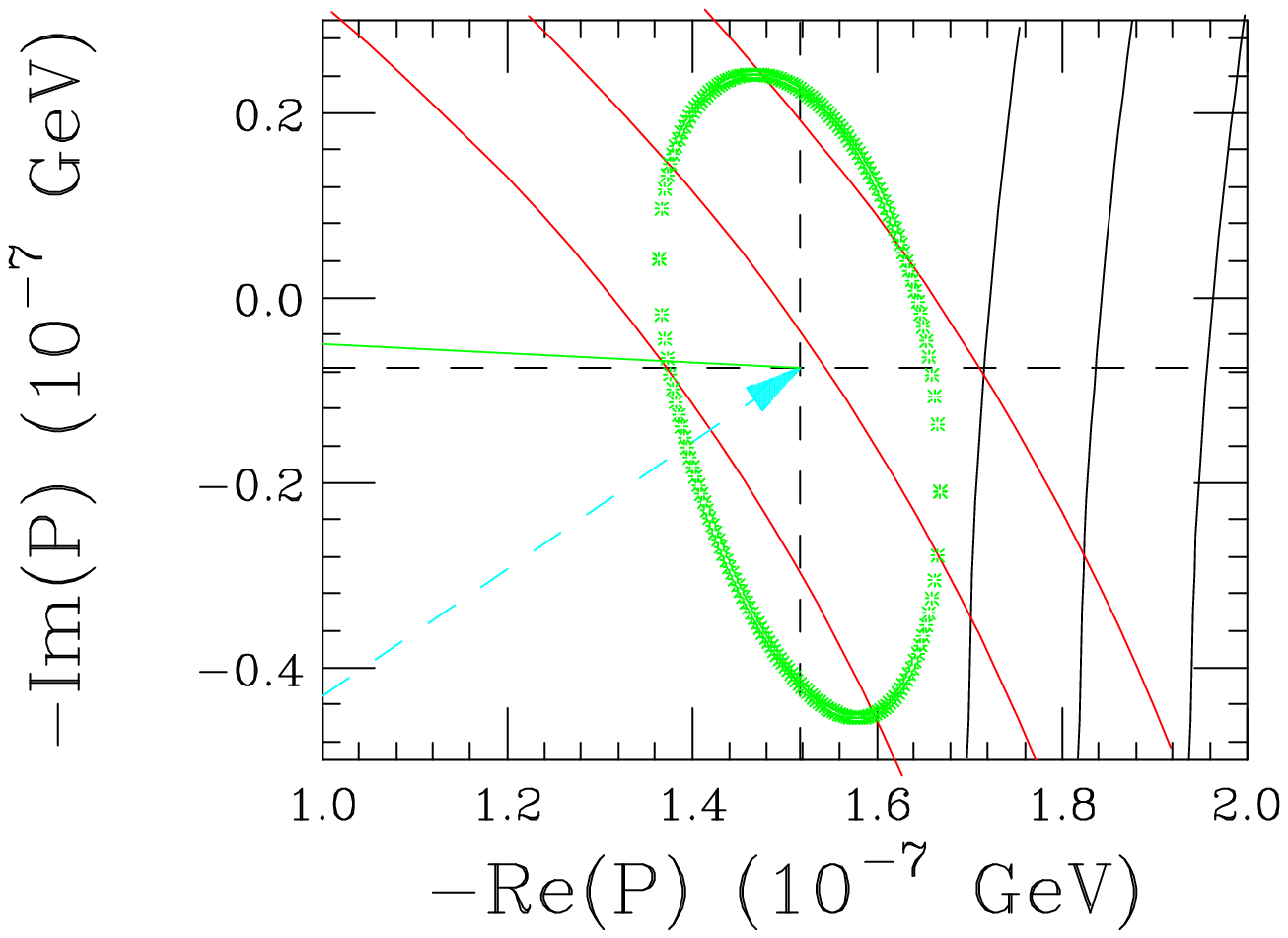}
\caption{Construction to determine $P$.
The relative sign between the left-hand and (magnified) right-hand
panels is due to the fact that the vector $P$ points {\it toward}
the origin in the left-hand figure.
 \label{fig:pe}}
\end{center}
\end{figure}

Using the extracted value of $P + PA$ we apply a similar construction technique
to extract $P$. The relevant parts of the tree-level amplitudes that determine
centers of the circles are as follows (in units of $10^{-7}$ GeV):
\bea
- (P + PA) &=& - 0.44 - 1.41~i~, 
\label{P+PA}\\
\lambda\,(T_K - A_{D^+}) &=& ~~8.37 - 0.58~i~, \\
- \lambda\,(T_\pi - A) &=& - 6.40 + 0.72~i~, \\
\lambda\,(C + A) &=& -4.51 - 1.85~i~.
\label{C+A}
\eea
The relevant experimental rates \cite{Bhattacharya:2009ps, Amsler:2008zzb} lead
to amplitudes once again determining the radii of the circles as follows
(in units of $10^{-7}$ GeV):
\bea
|\ca(D^0\to K^0\ok)| &=& 2.39 \pm 0.14,\\
|\ca(D^+\to K^+\ok)| &=& 6.55 \pm 0.12,\\
|\ca(D^+_s\to K^0\pi^+)| &=& 5.94 \pm 0.32,\\
\s\,|\ca(D^+_s\to K^+\pi^0)| &=& 2.94 \pm 0.55~.
\eea
$\chi^2$--minimization gives us
\beq\label{P}
P = [(- 1.52\pm0.15) + (0.08^{+0.38}_{-0.32})~i] \times 10^{-7}~{\rm GeV}~;
~~~~~~~\chi^2/{\rm d.o.f.} = 54/2 = 27~.
\eeq
The construction and the corresponding 68\% error ellipse ($\Delta \chi^2
= 2.3$) are shown in Fig.\ \ref{fig:pe}. In Table \ref{tab:pen} we quote the
representations and compare the experimental and fit amplitudes.
Using the extracted values of $P$ and $P + PA$ we find
\beq\label{PA}
PA = [(1.95\pm0.38) + (1.34\pm0.71)~i] \times 10^{-7}~{\rm GeV}~.
\eeq
We recall the two-fold ambiguity permitting amplitudes which are complex 
conjugates of (\ref{P+PA})-(\ref{C+A}) and (\ref{P})-(\ref{PA}).

The poor $\chi^2$ in this fit is driven primarily by the large contribution
from the $D_s^+ \to \pi^+ K^0$ amplitude.  It is quite possible that our
description of SU(3) breaking in this quantity is imperfect.  In any case, the
large experimental errors on the SCS decays of $D_s$ to two pseudoscalar mesons
will hinder the study of CP-violating asymmetries in their decays for some
time to come, so we shall not be greatly concerned with such decays for the
present.

\section{Description and prediction of observed direct CP asymmetries}

We now consider the effects of an additional phenomenological-penguin amplitude
$P_b$, the weak phase of which differs from the weak phase of $P$ and $PA$ by
the CKM-angle $\gamma$.  (The subscript $b$ refers to a $b$ quark in the
intermediate quark loop in Fig.\ \ref{fig:P}.)  In Table \ref{tab:phpen} we
summarize the amplitudes for SCS processes obtained in the previous section,
and extend the amplitude representations to include $P_b$.  The quantities
$\phi_T^f = {\rm Arg}[T_f]$ denote the strong phases of the non-$P_b$
amplitudes with respect to $T$.  (The amplitudes $T_f$ include factors
$\pm \lambda$.)

\begin{table}[h]
\caption{Fit amplitudes for SCS charmed meson decays including $P$ and $PA$,
and their representations including $P_b$. \label{tab:phpen}}
\begin{center}
\begin{tabular}{c c c} \hline \hline
Decay & Amplitude      & $\phi_T^f = {\rm Arg}[T_f]$ \\
 mode & representation & (degrees) \\ \hline
$D^0\to\pi^+\pi^-$ &$-\lambda\,(T_\pi + E) + (P + PA) + P_b$     &--158.5\\
$D^0\to  K^+  K^-$ &~$\lambda\,(T_K + E) + (P + PA) + P_b$       &32.5\\
$D^0\to\pi^0\pi^0$ &$-\lambda\,(C - E)/\s - (P + PA)/\s - P_b/\s$&60.0\\ \hline
$D^+\to\pi^+\pi^0$ &$-\lambda\,(T_\pi + C)/\s$                  &126.3\\ \hline
$D^0\to K^0\ok$    &$-(P + PA) + P$                              &--145.6\\
$D^+\to K^+\ok$    &~$\lambda\,(T_K - A_{D^+}) + P + P_b$        &--4.2\\
$D^+_s\to\pi^+ K^0$&$-\lambda\,(T_\pi - A) + P + P_b$            &174.3\\
$D^+_s\to\pi^0 K^+$&$-\lambda\,(C + A)/\s - P/\s - P_b/\s$ &16.4\\ \hline\hline
\end{tabular}
\end{center}
\end{table}

In general, the amplitude for $D \to f$ may be written as follows:
\beq
\ca(D\to f) = |T_f|\,e^{\i\,\phi^f_T}\,\left(1 + r_f\,e^{\i\,(\gamma + \phi^f)}\right),
\eeq
where $T_f$ represents terms that have the same weak phase as the tree-level
terms contributing to that amplitude, $\phi_T^f$ represents the strong phase of
$T_f$, $r_f$ represents the ratio of the magnitude of the CP-violating penguin
contribution to that of $T_f$, $\gamma$ represents the weak phase of the
CP-violating penguin (it is the same as the CKM angle), and $\phi^f$ is the
strong phase of the CP-violating penguin relative to $T_f$. Let us take the
example of the process $D^0\to\pi^+\pi^-$ for clarity. Then
\bea
T_{\pi^+\pi^-} &=& -\lambda\,(T_\pi + E) + (P + PA)~, \\
\phi^{\pi^+\pi^-}_T &=& {\rm Arg}[T_{\pi^+\pi^-}]~, \\
r_{\pi^+\pi^-} &=&\frac{|P_b|}{|T_{\pi^+\pi^-}|}~, \\
\phi^{\pi^+\pi^-} &=& {\rm Arg}[P_b] - \phi^{\pi^+\pi^-}_T - \gamma~.
\eea
The amplitude for $\ovD\to\of$ may be written as follows:
\beq
\ca(\ovD\to\of) = |T_f|\,e^{\i\,\phi^f_T}\,\left(1 + r_f\,e^{\i\,(-\gamma +
\phi^f)}\right)~.
\eeq
For a two-body decay, the rate is proportional to the absolute square of the
amplitude.  Thus, one may now define a CP asymmetry as follows:
\bea\label{Acp}
A_{CP}(f) &=& \frac{\G(D\to f) - \G(\ovD\to\of)}{\G(D\to f) + \G(\ovD\to\of)}
\nonumber\\
&=&- \frac{2\,r_f\,\sin\gamma\sin\phi^f}{1 + r^2_f + 2\,r_f\,\cos\gamma \cos
\phi^f} \nonumber\\
&=& - \frac{2\,p\,|T_f|\,\sin\gamma\sin(\delta - \phi^f_T)}{|T_f|^2 + p^2 +
2\,p\,|T_f|\, \cos\gamma\cos(\delta - \phi^f_T)},
\eea
where in the final step we have used $P_b = p\,e^{\i(\delta + \gamma)}$.

\begin{figure}
\begin{center}
\includegraphics[width=0.7\textwidth]{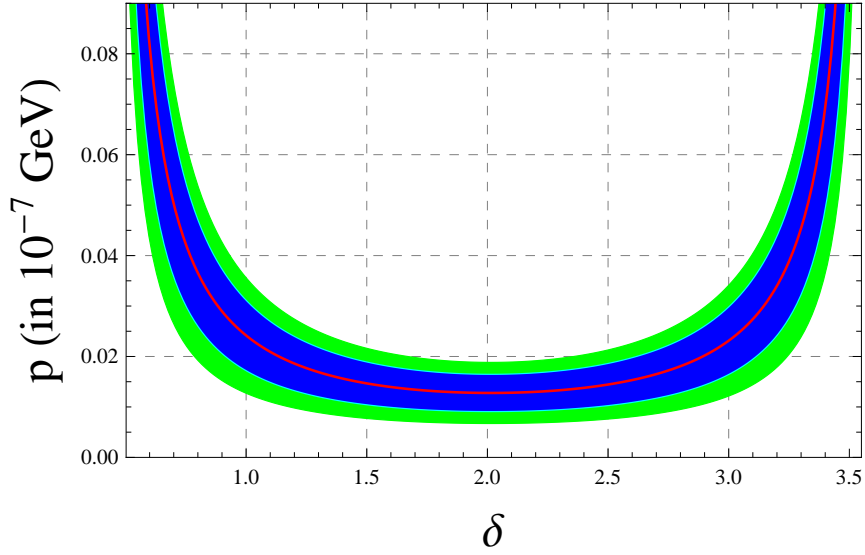}
\caption{$p$ and $\delta$ allowed by the measured range of $\Delta A_{CP}$. The
(red) line represents the central value, while inner (blue) and outer (green)
bands respectively represent 68\% confidence level (1$\sigma$) and 90\%
confidence level (1.64$\sigma$) regions based on error in $\Delta A_{CP}$.
\label{fig:pb}}
\end{center}
\end{figure}

The LHCb result (\ref{eqn:LHCb}) \cite{Aaij:2011in} may be used as a constraint
on the magnitude and strong phase of the CP-violating penguin $P_b$.
[Note added in proof:  To lowest order in $p$, all asymmetries $A_{CP}(f)$ 
depend on the combination $p \sin \gamma$.  Thus, if we impose the $\Delta
A_{CP}$ constraint, our predictions for other asymmetries are the same for any
weak phase of $P_b$ as long as effects of higher order in $p$ are negligible.
We thank N. Deshpande for a question leading to this result.]
We use the following relationships:
\bea\label{ACPkk}
A_{CP}(K^+K^-) &=& - \frac{2\,p\,|T_{K^+K^-}|\,\sin\gamma
\sin(\delta-\phi^{K^+K^-}_T)}{|T_{K^+K^-}|^2+p^2+2\,p\,|T_{K^+K^-}|\,
\cos\gamma\cos(\delta-\phi^{K^+K^-}_T)}~,\\
\label{ACPpipi}
A_{CP}(\pi^+\pi^-) &=& - \frac{2\,p\,|T_{\pi^+\pi^-}|\,\sin\gamma\sin(\delta
- \phi^{\pi^+\pi^-}_T)}{|T_{\pi^+\pi^-}|^2 + p^2 + 2\,p\,|T_{\pi^+\pi^-}|\,\cos
\gamma\cos(\delta - \phi^{\pi^+\pi^-}_T)}
\eea
We may use the theory fit results quoted in Table \ref{tab:pen} for $|T_f|$.
The strong phase $\phi^f_T$ can be taken from the results quoted in Table
\ref{tab:phpen}. 
The CKM angle $\gamma$ may be taken to be $77^\circ$
\cite{Nakamura:2010zzi}. Corresponding to each value of $\delta$ allowed by the
$\Delta A_{CP}$ constraint, one may extract the allowed values of $p$. In
addition we expect $|P_b| < |T_f|$ ($r_f < 1$), which in turn restricts us to
small values of $p$. In Fig.\ \ref{fig:pb} we plot the allowed values of $p$
as a function of $\delta$ using Eqs. (\ref{ACPkk}) and (\ref{ACPpipi}). 
For a wide range of $\delta$, a penguin amplitude of magnitude $0.01 \times
10^{-7}$ GeV, or ${\cal O}(0.1\%)$ of the $D^0 \to K^+ K^-$ amplitude, is
sufficient to account for the observed value of $\Delta A_{CP}$.  This is in
accord with a conclusion reached in Ref.\ \cite{Brod:2011re}.
%
\begin{figure}
\begin{center}
\includegraphics[width=0.7\textwidth]{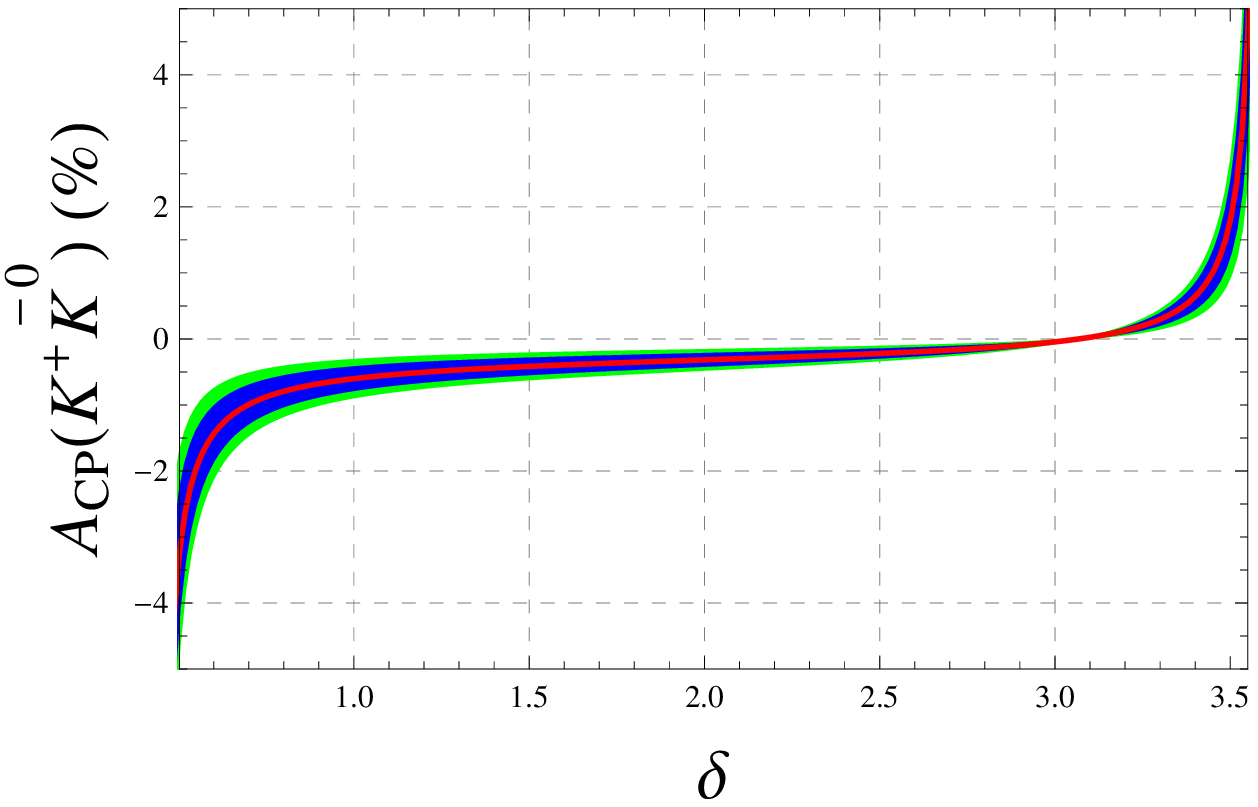}
\caption{$A_{CP}(K^+\ok)$ as a function of the allowed values of $\delta$. The
(red) line represents the central value, while inner (blue) and outer (green)
bands respectively represent 68\% confidence level (1$\sigma$) and 90\%
confidence level (1.64$\sigma$) regions based on error in $\Delta A_{CP}$.
\label{fig:KpK0b}}
\end{center}
\end{figure}
\begin{figure}
\begin{center}
\includegraphics[height=2.85in]{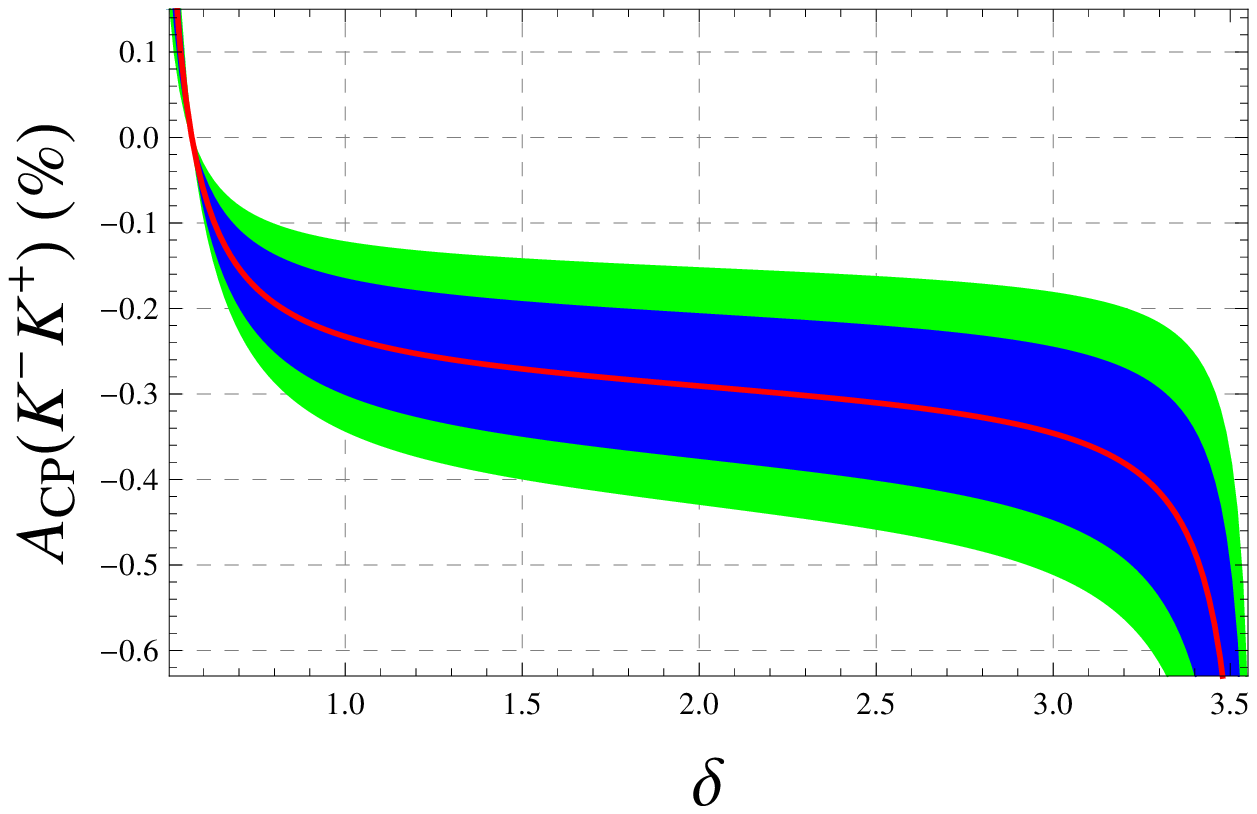} \hspace{0.5cm}
\includegraphics[height=2.85in]{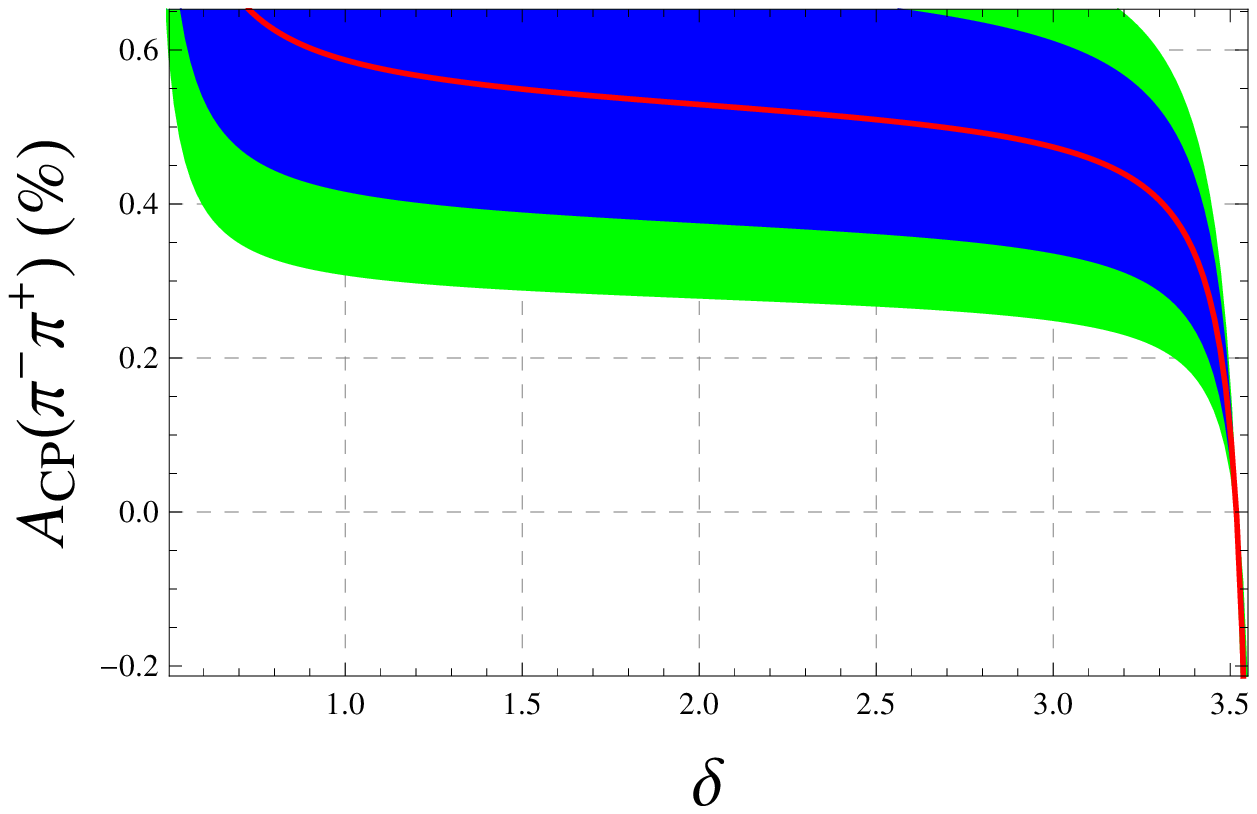}
\includegraphics[height=2.85in]{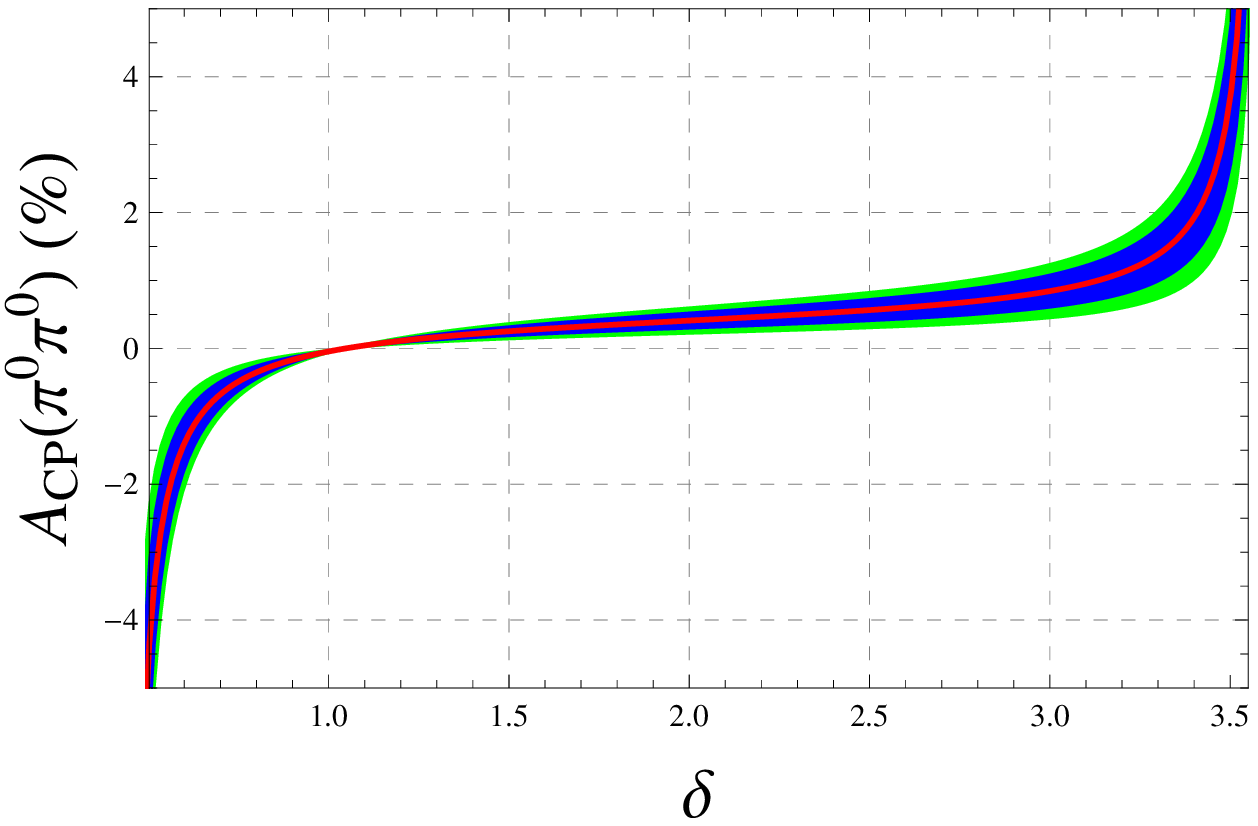}

\caption{$A_{CP}$ as a function of the allowed values of $\delta$. The
(red) line represents the central value, while inner (blue) and outer (green)
bands respectively represent 68\% confidence level (1$\sigma$) and 90\%
confidence level (1.64$\sigma$) regions based on error in $\Delta A_{CP}$.
\label{fig:ACP}}
\end{center}
\end{figure}

The constraint on $p$ as a function of $\delta$ now allows us to predict CP
asymmetries in other channels such as $D^+\to K^+ \ok$ as a function of the
angle $\delta$. In Fig.\ \ref{fig:KpK0b} we plot $A_{CP}(K^+\ok)$ as a
function of $\delta$.
Values of $p$ are plotted only for the range of $\delta$ consistent with
the limits (\ref{eqn:limits}), which will be specified shortly.  In Fig.\
\ref{fig:ACP} we plot $A_{CP}$ for the final states $K^+K^-$, $\pi^+ \pi^-$,
and $\pi^0\pi^0$.  The limits (\ref{eqn:limits}) imply the following allowed
ranges of $\delta$:
\beq
A_{CP}(K^+K^-) \Rightarrow 0.50 \le \delta \le 3.57 ~,~~
A_{CP}(\pi^+\pi^-) \Rightarrow 0.49 \le \delta \le 3.57.
\eeq
Figs.\ \ref{fig:pb} and \ref{fig:KpK0b} are plotted only for values of $\delta$
consistent with both these limits.  Note the correlation between the CP
asymmetries in the channels $D^0 \to \pi^0 \pi^0$ and $D^+ \to K^+ \ok$.
More precise measurements of the individual asymmetries in $D^0 \to \pi^+
\pi^-$ and $D^0 \to K^+ K^-$ can help to pin down the unknown strong phase
$\delta$.

As mentioned in the preceding section, all the contributions to $T_f$ listed 
in Table \ref{tab:phpen} involve an ambiguity due to complex conjugation. 
Thus, the phase $\phi^f_T$ has a sign ambiguity, $\phi^f_T\to -\phi^f_T$, 
which is common to all final states $f$. The CP asymmetry (\ref{Acp}) is 
approximately invariant under a joint transformation,
\beq
\phi^f_T \to -\phi^f_T~,~~~~\delta \to \pi - \delta~,
\eeq
neglecting a very small contribution to the asymmetry quadratic in $p/|T_f|$.
Thus, while  plots similar to Figs.\ \ref{fig:pb}, \ref{fig:KpK0b} and
\ref{fig:ACP} may be plotted with $\delta \to \pi-\delta$, the correlations
between asymmetries in different decay modes are invariant under this
redefinition. 

We have left out $D^+_s$ decay asymmetries
since the corresponding branching ratios have large fractional
errors. The process $D^+\to\pi^+ \pi^0$ does not depend on $P_b$
in the isospin symmetry limit, and therefore its CP asymmetry is zero
at this high level of approximation. The CP asymmetry in
$D^0\to K^0\ok$ depends only on a penguin annihilation diagram as
there are no $u$ quarks in the final state.
If it is found to be non-zero, our discussion must be expanded to include
the possibility of CP violation due to interference between
a $(PA)_b$ amplitude involving a $b$ quark in the loop and an SU(3)
breaking term in $E$.

\section{Discussion and summary}

The observation by the LHCb Collaboration of a difference between the
CP-violating asymmetries in $D^0 \to K^+ K^-$ and $D^0 \to \pi^+ \pi^-$
likely implies observable asymmetries in other decays of charmed mesons
to pairs of pseudoscalar mesons.  The present description of that difference
assumes that a penguin amplitude with an intermediate $b$ quark, normally
thought to provide a contribution below the
observed effect, is amplified by CP-conserving physics (e.g.,
unforeseen QCD effects) to an extent which can account for the asymmetry.  In
that case several direct CP asymmetries are predicted as functions of a single
strong phase difference $\delta$.  These include asymmetries in the individual
decays $D^0 \to K^+ K^-$ and $D^0 \to \pi^+ \pi^-$, as well as $D^0 \to \pi^0
\pi^0$ and $D^+ \to K^+ \ok$.  These asymmetries are typically of order (a few)
$\times 10^{-3}$, and the latter two are correlated with one another.
Experimental limits (\ref{eqn:limits}) on the direct CP
asymmetries in $D^0 \to K^+ K^-$ and $D^0 \to \pi^+ \pi^-$
\cite{Aaltonen:2011se} provide constraints on $\delta$.  The observed asymmetry
\cite{Link:2001zj} $A_{CP}(D^+ \to K^+ \ok) = (7.1 \pm 6.1 \pm 1.2)\%$
carries far too large an uncertainty at present to test its prediction.
[Note added in proof:  (1) The CDF Collaboration has now reported a value of
$\Delta A_{CP} = (-0.62 \pm 0.21 \pm 0.10)\%$ \cite{CDFDA}.  (2) We thank
Anze Zupanc for reminding us that the Belle Collaboration has reported the
much more precise value $A_{CP}(D^+ \to K^+ \ok) = (-0.16\pm0.58\pm0.25)\%$
\cite{Ko:2010ng}.]

In Fig.\ \ref{fig:ACP}, while $A_{CP}(K^+K^-)$ and $A_{CP}(\pi^+\pi^-)$
are predicted to have opposite signs for a wide range of $\delta$, their
relative {\it magnitudes} provide information about $\delta$, with the ratio
$|A_{CP}(\pi^+\pi^-)/A_{CP}(K^+K^-)|$ exceeding 1 for the mid-range
of $\delta$ and behaving as a decreasing function of $\delta$. Thus, better
measurements of these individual asymmetries will enable improved
predictions of asymmetries such as $A_{CP}(K^+ \ok)$ and $A_{CP}(\pi^0 \pi^0)$.
We look forward to improvement of many of these determinations.

\section*{Acknowledgements}

B. B. would like to acknowledge the hospitality of the Particle Theory Group,
University of Chicago during his stay in Chicago.  He also thanks David London
for useful discussions.  This work was supported in part by the United States
Department of Energy through Grant No.\ DE FG02 90ER40560.

\end{document}